\documentclass[12pt] {article}

 \font\gotb eufm10 scaled \magstep1
\newcommand{\bb}{\bibitem}
\newcommand{\cc}{\cite}
\newcommand{\vp}{\varphi}
\newcommand{\sss}{\sigma}
\newcommand{\al}{\alpha}
\newcommand{\Om}{\Omega}
\newcommand{\om}{\omega}
\newcommand{\lt}{\left}
\newcommand{\rt}{\right}
\newcommand{\lll}{\lambda}
\newcommand{\F}{{\cal F}}
\newcommand{\Ss}{\hat S^2_}
\newcommand{\RR}{\bar R}

\newcommand{\am}{\hat a^-}
\newcommand{\ap}{\hat a^+}
\newcommand{\R}{\hat R}
\newcommand{\s}{\hat S}
\newcommand{\K}{\hat K}
\newcommand{\I}{\hat I}
\newcommand{\Q}{\hat Q}
\newcommand{\J}{\hat J}
\newcommand{\Aa}{\bar A}
\newcommand{\A}{\hat A}
\newcommand{\B}{\hat B}
\newcommand{\U}{\hat U}
\newcommand{\p}{\hat p}
\newcommand{\PP}{\hat P}
\newcommand{\po}{\Psi_0}
\newcommand{\aA}{\tilde A}
\newcommand{\HH}{\hat H}
\newcommand{\tvp}{\tilde \vp}
\newcommand{\AAA}{\hbox{\gotb A}}
\newcommand{\AC}{$\hbox{\gotb A}_{cl}$}
\newcommand{\QQ}{\hbox{\gotb Q}}
\newcommand {\QQQ}{\hbox {\gotb Q}_{\xi}}
\newcommand {\qqq}{\hbox {\gotb Q}_{\xi'}}
\newcommand {\vx}{\vp_{\xi}}
\newcommand {\vpx}{\vx(\A)}
\newcommand{\BB}{\hbox{\gotb B}}
\newcommand{\bea}{\begin{eqnarray} \label}
\newcommand{\eeq}{\end{equation}}
\newcommand{\beq}{\begin{equation} \label}
\newcommand{\eea}{\end{eqnarray}}
\newcommand{\nn}{\\ \nonumber}
\newcommand{\rr}[1]{(\ref{#1})}

\author{D.A. Slavnov}

\title {Quantum mechanics without quantum logic}

   \date {}

\begin {document}

   \maketitle
   \begin{center} {\it  Department of Physics, Moscow State
University,\\  Moscow 119899, Russia. E-mail:
slavnov@goa.bog.msu.ru }
 \end{center}

 \begin {abstract}
We describe a scheme of quantum mechanics in which the Hilbert
space and linear operators are only secondary structures of the
theory. As primary structures we consider observables, elements of
noncommutative algebra, and the physical states, the nonlinear
functionals on this algebra, which associate with results of
single measurement. We show  that in such scheme the mathematical
apparatus of the standard quantum mechanics does not contradict  a
hypothesis on existence of an objective local reality,  a
principle of a causality and Kolmogorovian probability theory.
\end {abstract}

\section {Introduction}

Quantum field theory has achieved significant successes in the
last decades. These successes are related to creation of
non-Abelian (noncommutative) gauge models. The newest quark
physics has arisen on their basis.  Abelian gauge model,  quantum
electrodynamics, was known for a long time. Transition to
non-Abelian models became a qualitative leap in development of
quantum field theory.

At the same time, this transition has not caused any essential
revision of the basic concepts of quantum field theory.
Especially, it has not demanded any changes in logic and
mathematics.

In present paper the idea is carried out that transition from
classical  to quantum  physics is similar to transition from
Abelian to non-Abelian  gauge models. Of course, the quantum
physics is qualitatively new theory. However,  for successful
development of the quantum theory it is completely not necessary
to refuse the main concepts of the classical theory: the formal
logic, the classical probability theory, the principle of
causality, idea on the objective physical reality.

The base notions of the modern standard quantum mechanics are the
Hilbert space and linear operators in this space. Von
Neumann~\cc{von} mathematically precisely has formulated quantum
mechanics on the basis these concepts. The matrix mechanics of
Heisenberg and the wave mechanics of  Schr\"odinger are concrete
realization of the von Neumann's abstract method.

The formalism of the Hilbert space became the mathematical basis
of those tremendous successes  which were achieved by quantum
mechanics. However, these successes have also a reverse side.
There is some worship of the Hilbert space. Physicists have ceased
to pay attention that the Hilbert space rather specific
mathematical object. It has appeared a excellent basis for
calculation of expectation values of observable quantities and
their probabilistic distributions. At the same time, it is
completely not self-evident that observables are operators in the
Hilbert space.

Attempts to use the formalism of the Hilbert space for the
description of the single physical phenomena not so are
successful. Reasonings which are used in this case,  frequently
appear not indisputable.

So von Neumann~\cc{von}  resorts to rather doubtful idea on
"internal I"  to coordinate the concept of the Hilbert space to
the results of single measurements. Abandonment  of the causality
principle also is hard to perceive.

The same concerns the idea on the determinative influence of the
observer onto quantum-mechanical processes.  In this respect
quantum mechanics is a unique subdiscipline of physics and  of the
science in general. The idea (see for example,~\cc{men}) is put
forward on the active role of conscious  in the quantum phenomena.

In attempt to give slightly more objective form to similar notion
Everett~\cc {ever} has put forward rather exotic idea on existence
in the nature of set of the parallel worlds. Happy-go-lucky the
observer appears at each  measurement in any one of these worlds.
This idea  has got very many supporters despite  all extravagance.

Probably, this specifies that though the Hilbert space rather
useful mathematical object, its base role is completely not
indisputable. It is not a new idea  that the Hilbert space and
linear operators are not primary elements of the quantum theory.
Namely this idea became a basis of the algebraic approach to
quantum field theory (see for example,~\cc{emch,hor,blot}).

 \section {Observables and states}

Base notion of  classical physics is "observable". This notion
seems self-evident and does not demand definition.  It is possible
to multiply them by real numbers, to sum up and multiply together.
In other words, they form a real algebra \AC.

The elements $ \A $ of this algebra are the latent form of
observable variables. The explicit form of an observable should be
some number. It means that  the explicit form of an observable
corresponds to the value of some functional $ \vp(\A)=A $ ($ A $
is a real number), defined on the algebra \AC.

Physically, the latent form of the observable $ \A $ becomes
explicit as a result of measurement. This means that the
functional $ \vp(\A) $ describes a measurement result of the
observable $ \A $.

Experiment shows  that the sum and the product of measurement
results correspond to the sum and the product of observables:
 \beq{0}
  \A_1 +\A_2 \to A_1+A_2, \qquad \A_1\A_2 \to A_1A_2. \eeq

In this connection further there will be useful a following
definition~\cc{rud}.

Let \BB{} be a real  commutative algebra and $ \tvp $ be a linear
functional on~\BB. If
 $$ \tvp(\B_1\B_2) = \tvp(\B_1)
\tvp(\B_2)\mbox{ for all }\B_1\in\BB \mbox{ and } \B_2\in\BB, $$
then the functional $ \tvp $ is called a real  homomorphism on
algebra~\BB.

Now we can introduce the second base notion of  classical physics
--- a state of  the object under consideration.

The state is a real homomorphism on the algebra of observables.
The result of any measurement of the classical object is
determined by its state.

In principle, in classical physics it is possible to measure
observables in any combinations. Always it is possible to pick up
such system of measuring devices that measurement results of
several observables will not depend on sequence in what
observables are measured. For example, if we carry out
measurement of an observable $ \A $, then an observable $ \B $,
then again an observable $ \A $ and an observable $ \B $ results
of repeated measurements of observables  coincide with results of
primary measurements. We  name the corresponding measuring
instruments compatible devices.

Let us pass to discussion of the situation in quantum physics. In
quantum physics  it seems also natural to accept  "observable" as
base notion. Quantum observables also possess algebraic
properties.

However quantum measurements significantly differ from classical
one. There are systems of compatible measuring devices not for any
observables. Accordingly, in quantum physics observables are
subdivided into compatible (simultaneously measurable) and
incompatible (additional). There are systems of compatible
measuring devices for compatible observables. Such systems of
devices do not exist for incompatible observables. As it is told
in paper by Zeilinger~\cc{zeil}: "Quantum complimentarity then is
simply expression of the fact that in order to measure quantities,
we  would have to use apparatuses which mutually exclude each
other".

For incompatible quantum observables the measurement results
depend on sequence of measurement of these observables. This fact
leaves traces on rules of multiplication of observables. Two ways
are applied. The first way is used in  so-called Jordan
algebra~\cc{jord,emch}. In this case, observables form real
commutative algebra, but the operation of multiplication is not
associative. Use of Jordan algebra  has not led to to appreciable
successes in the quantum theory.

The standard quantum mechanics is based on the other method of
multiplication of observables. In this case, operation of
multiplication is associative, but noncommutative. Besides product
of two observables not necessarily is an observable, i.e.
observables do not form algebra.

In order to  use the advanced mathematical apparatus of algebras
full-scale, it is convenient to leave for framework of the
directly observable  variables and to consider its complex
combinations.  Hereinafter, we call these combinations  the
dynamic variables.

Having in view of told above, we  accept

\

 {\it  Postulate 1:}

{\it Dynamic variables correspond to elements of an involutive,
associative, and (in general) noncommutative algebra~\AAA,
satisfying the following conditions: for each element $
\R\in\AAA$, there exists a Hermitian element $\A $ $ (\A^*=\A) $
such that $ \R^*\R =\A^2 $, and if $\R^*\R=0$, then $ \R=0 $}.

\

We assume that the algebra has a unit element~$\I $ and that
Hermitian elements of algebra~\AAA{} correspond to observable
variables.  We let~$\AAA_+ $ denote the set of these elements.

In the standard quantum mechanics this postulate is accepted in
considerably stronger form. It is supposed that dynamic  variables
$ \R $ are linear operators in the Hilbert space.

Postulate 2 directly follows  from quantum measurements

 \

{\it  Postulate 2:}

{\it  Mutually commuting elements of the set $ \AAA_+ $ correspond
to compatible (simultaneously measurable) observables.}

\

Because of Postulate 2, commutative subalgebras of the algebra
\AAA{} have an  important role in the further analysis.
(see~\cc{slav1}).

For the further it is useful to recollect definition of the
spectrum $ \sss (\A; \AAA) $ of an element $ \A $ in the
algebras~\AAA. The number $\lll $ is a point of the spectrum of
the element~$ \A $ if and only if the element $ \A-\lll\I $ does
not have an inverse element in the algebra~\AAA. Generally, an
element may have different spectra in an algebra and its
subalgebra.  However, if the subalgebra~\QQ {} is maximal, $ \sss
(\Q; \QQ) = \sss (\Q; \AAA) $ for any $ \Q\in\QQ $ (see for
example,~\cc{rud}).

The Hermitian elements of the algebra~\AAA{} are the latent form
of the observable variables. The same as in a classical case the
latent form of an observable $ \A $ becomes explicit as a result
measurements. Only mutually commuting observables can be are
measured in individual experiment. Experiment shows that the same
relations~\rr{0} are carried out  for  such observables like for
classical observables.

Generalizing definition of a state in classical physics, we accept
the central postulate of the proposed approach (see~\cc{slav2}).

\

 {\it  Postulate 3:}

 {\it The result of the observation which we carry out on the quantum system is determined by a
physical state of this system. The physical state is described by
a functional $ \vp (\A) $ (generally, multivalued), with
$\A\in\AAA_+ $, whose restriction $ \vpx $ to each
subalgebra~$\QQQ $ is single-valued and is a real homomorphism
($\vpx=A $ is a real number).}

 \

The functionals  $\vpx$  can be shown to have the  following
properties~\cc{rud}:

  \bea{1}
&1)& \vx(0)=0; \nn{} &2)& \vx(\I)=1; \nn{} &3)& \vx(\A^2)\ge 0;
\nn{}  &4)& \mbox{ if } \lll=\vx(\A), \mbox{ then }
\lll\in\sss(\A;\QQQ); \nn{} &5)& \mbox{ if } \lll\in\sss(\A;\QQQ),
\mbox{ then } \lll=\vx(\A) \mbox{ for some } \vx(\A).
 \eea
  The corresponding properties of individual measurements are postulated
in the  standard quantum mechanics but are a consequence of the
third postulate here.

On the other hand, properties (2.4) and (2.5) allow to construct
all functionals $ \vp $, appearing in the Postulate 3. Clearly
that for construction of the functional $ \vp $ sufficiently to
construct all its restriction $ \vx $ on subalgebras $ \QQQ $. In
its turn, each functional $ \vx $ can be constructed as follows.
In each subalgebra $ \QQQ $ it is necessary to choose arbitrarily
system $G(\QQQ) $ independent generators. Further we require $\vx$
to be a certain mapping $G(\QQQ) $ to a real number set (allowable
points of the spectra for the corresponding elements of the set
$G(\QQQ) $). On the other elements of $ \QQQ $, the functional $
\vx $ is constructed by linearity and multiplicativity. Sorting
out all possible mappings of the set $G (\QQQ) $ into points of
the spectrum, we construct all functionals $ \vx $.

On other subalgebra $ \qqq $ the functional $ \vp_{\xi'} $ is
constructed similarly. It is clearly that this procedure is always
possible if functionals $ \vx $ and $ \vp_{\xi'} $ are constructed
independently.

Thus, the set of physical states (functionals $ \vp $) is
completely defined by the algebra \AAA {} (set of its maximal real
commutative subalgebras and spectra of these subalgebras). It is
clearly that these functionals, generally, are multivalued.
Moreover, it is possible to show~\cc{ksp} that there are algebras
having physical sense for which it is impossible to construct the
single-valued functional

However, it is always possible to construct a functional $ \vp $
that is single-valued on all observables belonging any preset
subalgebra $ \QQQ $. For this, suffice it to assign number one
(set $ \xi=1 $) to the subalgebra $ \QQQ $  and define the
restriction $ \vp_1 $ of $ \vp $ to $ \QQ_1 $ as follows. Let
$G(\QQ_1) $ be a set of generators of $ \QQ_1 $. We define the
restriction $ \vp_1 $ to be some mapping of $G (\QQ_1) $ to a real
number set $S_1 $. We next choose another subalgebra $ \QQ_2 $.
With $ \QQ_1\cap \QQ_2\equiv \QQ_{12} \ne \emptyset $, we first
construct a set of generators $ G_{12} $ of $ \QQ_{12} $, and then
supplement it with the set $G_{21} $ to the complete set of
generators of $ \QQ_2 $. If $ \A\in G_{12} $, then $ \vp_2 (\A) =
\vp_1 (\A) $. If $ \A\in G_{21} $, then the functional $ \vp_2 $
is defined such that it is a mapping of $ G_{21} $ to some
allowable set of points in the spectra of the corresponding
elements of the algebra $ \QQ_2 $. We must next exhaust all
subalgebras $ \QQ_i $ (of type $ \QQQ $) that have nonempty
intersections with $ \QQ_1 $. To construct the restriction $ \vp_i
$ of $ \vp $ to each $ \QQ_i $, suffice it to use the recipe used
for $ \vp_2 $. By construction, such a functional $ \vp $ is
single-valued on all elements belonging to $ \QQ_1 $. Different
subalgebras $ \QQ_i $ can have common elements that do not belong
to $ \QQ_1 $. On these elements, the functional $ \vp $ can be
multivalued.

Physically the multivaluedness of the functional $ \vp $ can be
justified as follows. The result of observation may depend not
only on an observable quantum object, but also on properties of
the measuring device used for observation. A typical measuring
device consists of an analyzer and a detector. The analyzer is a
device with one input and several output channels. As an example,
we consider the device measuring an observable $ \A $. For
simplicity, we assume that the spectrum of this observable is
discrete. Each output channel of the analyzer must then correspond
to a certain point of the spectrum. The detector registers the
output channel through which the quantum object leaves the
analyzer. The corresponding point of the spectrum is taken to be
the value of the observable $ \A $ registered by the measuring
device.

In general, the value not of one observable $ \A_i $ but of an
entire set of compatible observables can be registered in one
experiment. All these observables must belong to a single
subalgebra $ \QQQ $. Naturally the analyzer must be constructed
appropriately. The main technical requirement to this construction
consists in the following. Sorting of researched quantum objects
on values of one of the observables belonging~$ \QQQ $, should not
deform sorting on values of other observable belonging~$ \QQQ $.

Therefore, measuring devices can be are marked by an index $\xi $.
We say that the corresponding device belongs to the type $ \QQQ $.

Let now we want to measure value of the observable
$\A\in\QQQ\cap\qqq $. For this purpose we can use the device of
the type $ \QQQ $ or the device of the type $ \qqq $. These
devices have a different construction. Therefore, it is absolutely
not necessary that we  obtain the same result if we  use a
different devices for investigation of the same quantum object. It
just corresponds to what the functional describing the physical
state of the quantum object, is multivalued on the observable
$\A$.

We cannot use both devices in one experiment as subalgebras $\QQQ$
and $ \qqq $ contain incompatible observables. Therefore, in each
concrete experiment the measurement result is single-valued.
However, this result can depend not only on internal
characteristics of the observable object (its physical state), but
also on the characteristic (type) of the measuring device.
Accordingly, the functional $ \vp $ describes not the value of
each observable in a given physical state but the response of the
measuring device of defined type, to this physical state.

Here there is an essential difference between the proposed
approach and so-called PIV model described in the review~\cc{hom}.
In this model it is supposed that the value of each observable is
uniquely predetermined for the quantum object.

If the functional $ \vp $ is single-valued at a point $ \A $, we
say that the corresponding physical state $ \vp $ is stable on the
observable $ \A $. In this case, we can say that the observable
$\A $ has a definite value in the physical state $ \vp $ and this
value is the physical reality.

The commutative algebra \AAA{} has only one maximal real
commutative subalgebra. Therefore, in this case all measuring
devices belong to one type and all physical states are stable on
all observables.

It is natural to impose a condition of "minimality" on algebra
\AAA{} with the following physical sense. If we cannot distinguish
two observables by any experiment it is one element of algebra.
Therefore, we accept

\

{\it Postulate 4:}

{\it If $ \sup_{\vx}|\vx(\A_1-\A_2)| =0 $, then $ \A_1 = \A_2 $.}

\

We note that the standard quantum mechanics implies the stronger
assumption that the observables coincide if all their average
values coincide.

It is obvious that in case of commutative algebra, the set of the
physical states passes into  usual classical phase space, and a
separate functional $ \vp $ comes to a point in this space.
However, for noncommutative algebras the physical state differs
that is understood as a state in quantum physics. Further we use
the term "a quantum state" for latter.

Let us note that the physical state  can not be fixed uniquely in
quantum measurement. Really, the devices for measurement of
incompatible observables are incompatible. Therefore, we can
measure the observables belonging to one maximal commutative
subalgebra $ \QQQ $ in one experiment. As a result we establish
only values of the functional $ \vx $ which is a restriction of
the functional $ \vp $ (the physical state). In other respects the
functional $ \vp $ is uncertain. Repeated measurement with use of
different device does not help the situation. Because the quantum
object passes into a new physical state for which the values of
the functional found in the first experiment $ \vx $ is useless.

We say that the functionals  $ \vp $ are $ \vx $-equivalent if
they have the same restriction $ \vx $ on the subalgebra $ \QQQ $.
Thus, in quantum measurement we can ascertain only the class of
equivalence $ \{\vp \}_{\vx} $ to which the physical state
belongs. Because the functional $ \vp $ has continual set of
restrictions on various maximal commutative subalgebras the set $
\{\vp \}_{\vx} $ has a potency of the continuum. Only this class
of equivalence can claim a name "quantum state".

Actually there is one more restriction. Experiment shows that if
the quantum object is in the quantum state corresponding
$\{\vp\}_{\vx} $ the measurement result of the observable
$\A\in\QQQ $ does not depend on the type of used measuring device.
It means that the physical state should be stable on the
subalgebra $ \QQQ $. We call such class of equivalence the quantum
state and denote by $ \Psi (\,\cdot \,| \vx) $.

Strictly speaking, the above definition of the quantum state is
valid only for a physical system that does not contain identical
particles. Describing identical particles requires some
generalization of the definition of the quantum state.

The measuring instrument cannot distinguish which of the identical
particles entered it. Therefore, we slightly generalize the
definition of a quantum state. Let the physical system contain
identical particles. Let $ \{\vp \}_{\vx} $ be the set of $ \vx
$-equivalent functionals. We say that a functional $\vp'$ is
weakly $ \vx $-equivalent to the functionals $ \vp $ if its
restriction $ \vp'_{\xi'} $ on the subalgebra $ \qqq $ coincides
with  the restriction $ \vp_{\xi}$ of the functional $ \vp $ on
the subalgebra $ \QQQ $. The set $ \{\Q'\} $ must be an image of
the set $ \{\Q \} $ under a mapping such that the observables
corresponding to one of the identical particles are changed to the
respective observables corresponding to the other particle.

Hence, the definition of the quantum state $ \Psi (\,\cdot \,|
\vx) $ should refer to the set of all weakly $ \vx $-equivalent
functionals. Hereinafter, we let the symbol $\{\vp\}_{\vx} $
denote the set of a weakly $ \vx $-equivalent functionals.

\section {Probability theory and quantum ensemble}

Now the Kolmogorovian probability theory~\cc{kol} is the most
consistent and mathematically rigorous. It is usually considered
that it does not approach for the description of quantum systems.
In present paper the opposite opinion is protected: the
Kolmogorovian probability theory very well approaches for the
description of the quantum phenomena, it is necessary only to take
into account peculiarity of quantum measurements~\cc{slav3}.

We recall the foundations of Kolmogorov's theory probability (see,
for example~\cc{kol,nev}). The probability theory scheme is based
on the so-called probability space $ (\Om, \F, P) $.

The first component $ \Om $ is set (space) of the elementary
events $ \om $. The physical sense of the elementary events
specially is not stipulated, but it is considered that they are
mutually exclusive. One and only one elementary event is realized
in each test.

Besides the elementary event the concept of "event" (or "random
event") is introduced. Each event $F $ is identified with some
subset of set $ \Om $. It is supposed that we can ascertain, the
event is carried out or failed in a experiment under
consideration. Such assumption is not done about the elementary
event.

Collections of subsets of the set $ \Om $ (including the set $\Om$
and the empty set $ \emptyset $) are supplied with the structure
of Boolean algebras. Algebraic operations are: intersection of
subsets, joining of them, and complement with respect to $ \Om $.
A Boolean algebra, closed in respect of denumerable number of
operations of joining and intersection, is called a
$\sss$-algebra.

The second component of the probability space is some
$\sss$-algebra $ \F $. The set $ \Om $ in which the particular
$\sss $-algebra $ \F $ is chosen, refers to as measurable space.
Further on the measurability will play a key role.

Finally, the third component of the probability space is the
probability measure $P $. It represents a mapping of the algebra
$\F $ onto a set of real numbers satisfying the following
conditions for any countable set of nonintersecting subsets
$F_j\in \F $: a) $0\leq P (F) \leq 1 $ for all $F\in\F $,
$P(\Om)=1 $; b) $P (\sum_j F_j) = \sum_j P (F_j) $ Let us pay
attention that the probabilistic measure is defined only for the
events which are included in the algebra $ \F $. For the
elementary events the probability, generally speaking, is not
defined.

A real random quantity on $ \Om $ is defined as a mapping $X $ of
the set $ \Om $ onto the extended real straight line $ \RR =
[\infty, + \infty] $, $$ X (\om) =X \in \RR. $$

The set $ \RR $ is assumed to have the measurability property. The
Boolean algebra $ \F_R $ generated by the semi-open intervals
$(x_i, x_j] $, i.e., the $ \sss $-algebra that results from
applying the algebraic operations to all such intervals, can be
chosen as the $ \sss $-algebra in the set $ \RR $. Let $ \{\om: X
(\om) \in F_R \} $, where $F_R\in \F_R $ be the subset of
elementary events $ \om $ such that $X (\om) \in F_R $. The
subsets $F = \{\om: X (\om) \in F_R \} $ form the $ \sss$-algebra
$ \F $ in the space~$ \Om $.

We consider now the application of formulated main principles of
probability theory to a problem of quantum measurements. We
associate an elementary event with a physical state. Accordingly,
we associate the set of physical states of a quantum object with
the space $ \Om $. Further, we need to make this space measurable,
i.e. to choose certain $ \sss $-algebra $\F$. Here, a peculiarity
of quantum measurement, which has the name "principle of
complimentarity" in standard quantum mechanics, has crucial
importance. We can organize each individual experiment only in
such a way that compatible observables are measured in it. The
results of measurement can be random. That is, observables
correspond to the real random quantities in probability theory.

The main goal of a typical quantum experiment is to obtain the
probability distributions for one or another observable
quantities. We can obtain such distribution for certain collection
of compatible observables if an appropriate measuring device is
used. From the point of view of probability theory we choose
certain $ \sss $-algebra $ \F $, choosing the certain measuring
device. For example, let us use the device intended for
measurement of momentum of a particle. Let us suppose that we can
ascertain by means of this device that the momentum of particle
hits an interval $ (p_i, p_j] $. For definiteness we have taken a
semi-open interval though it is not necessary. Hit of momentum of
the particle in this or that interval is the event for the
measuring device, which we use. These events are elements of
certain $ \sss $-algebra. Thus, the probability space $ (\Om, \F,
P) $ is determined not only by the explored quantum object (by
collection of its physical states) but also by the measuring
device which we use.

Let us assume that we carry out some typical quantum experiment.
We have an ensemble of the quantum systems, belonging to a certain
{\it quantum} state. For example, the particles have spin 1/2 and
the spin projection on the $x $ axis equals 1/2. Let us
investigate the distribution of two incompatible observables (for
example, the spin projections on the directions forming angles $
\theta_1 $ and $ \theta_2 $ with regard to the $x $ axis). We
cannot measure both observables in one experiment. Therefore, we
should carry out two groups of experiments which use different
measuring devices. "Different" is classically distinct. In our
concrete case the devices should be oriented by various manners in
the space.

We can describe one group of experiments with the help of a
probability space $ (\Om, \F_1, P_1) $, another group with the
help of $ (\Om, \F_2, P_2) $. Although in both cases the space of
elementary events $ \Om $ is the same, the probability spaces are
different. Certain $ \sss $-algebras $ \F_1 $ and $ \F_2 $ are
introduced in these spaces to give them the property of
measurability.

Mathematically~\cc{nev}, a $ \sss $-algebra $ \F _ {12} $ that
include the algebra $ \F_1 $ as well as algebra $ \F_2 $ can be
formally constructed.

It is said that such an algebra is generated by the algebras
$\F_1$ and $ \F_2 $. In addition to the subsets $F^{(1)}_i\in\F_1$
and $F^{(2)}_j\in \F_2 $ of the set $ \Om $, it also contains all
possible intersections and unions of the subsets
$F^{(1)}_i\in\F_1$ and $F^{(2)}_j\in \F_2 $.

But this $ \sss $-algebra is unacceptable physically. Indeed, the
event $F_{ij} =F^{(1)}_i\cap F^{(2)}_j $ is an event in which the
values of two incompatible observables of one quantum object
belong to a strictly determined domain. For a quantum system, it
is impossible in principle to set up an experiment that could
distinguish such an event. Therefore, the probability concept does
not exist for such event. In other words, there is no probability
measure corresponding to the subset $F_{ij} $, and the
$\sss$-algebra $ \F_{12} $ cannot be used to construct the
probability space. This illustrates the following fundamental
point that should be kept in mind when applying the theory of
probability to quantum systems: not all mathematically possible
$\sss $-algebras are physically acceptable.

The probability definition implies numerous tests. These tests
must be performed under same conditions. This applies to the
object being tested as well as to the measuring instrument. It is
obvious that the microstates of either the object or the
instrument cannot be fully controlled. Therefore, the term "the
same conditions" should refer to some equivalence classes for the
states of the quantum object and the measuring instrument.

For a quantum object under study, such fixation is normally
realized by choosing a certain quantum state. For example, in the
case of spin particles, the particles with a certain spin
orientation are selected.

For the measuring instrument, we also must choose a definite
classical characteristic to be used to fix a certain equivalence
class. For example, the initial single beam of particles in the
instrument should split into a few well-separated beams
corresponding to different values of the spin projection on some
distinguished direction

Thus, what corresponds to an element of the measurable space $
(\Om, \F) $ in an experiment is the ensemble of quantum objects
(which can be in a definite quantum state) and a measuring
instrument of a certain type that allows registering an event of a
definite form. Each such instrument can distinguish events that
correspond to some set of compatible observables. As it was
already mentioned, the result of individual an measurement may
depend not only on intrinsical properties of the measured object
(the physical state), but also on the type of the measuring
device. In terms of the probability theory, this can be expressed
as follows for a quantum system, a random variable $X $ can be a
multivalued function of the elementary event $ \om $.

In the classical case, all observables are compatible.
Accordingly, all measuring instruments belong to one type;
therefore, the classical random quantity $X $ is a single-valued
function of $ \om $. We note that in the quantum case, if the
quantity $X $ is interpreted as a function on the measurable space
$ (\Om, \F) $ rather than the space $ \Om $, then this function is
single-valued.

All this motivates us to reconsider the interpretation of the
result obtained in~\cc{ksp}, where a no-go theorem was proved.
Essentially, the theorem states that there is no intrinsic
characteristic of a particle with spin 1 that unambiguously
predetermines the squares of the spin projections on three
mutually orthogonal directions.

The conditions of the Kochen-Specker theorem are not carried out
in the approach described in present paper. Really, used in
paper~\cc{ksp} the observables $ (\Ss x, \Ss y, \Ss z) $ are
compatible. The observables $ (\Ss {x}, \Ss {y'}, \Ss {z'}) $ are
also compatible. Here, the $x, y', z' $ directions are orthogonal
among themselves, but the $y, z $ directions are not parallel to
the $y', z' $ directions. The observables $ (\Ss y, \Ss z) $ are
not compatible with the observables $ (\Ss {y'}, \Ss {z'}) $. The
devices coordinated with the observables $ (\Ss x, \Ss y, \Ss z) $
and the $ (\Ss {x}, \Ss {y'}, \Ss {z'}) $ belong to different
types. Therefore, these devices not necessarily should give the
same result for square of spin projection on the $x $ direction.
It is impossible to carry out the experiment for check of this
statement, as we cannot use simultaneously two types of measuring
devices in one experiment.

Let us consider an ensemble of physical systems which are in the
quantum state $ \Psi (\, \cdot \, | \vx) $. We consider the
physical states $ \vp $ of these systems as an elementary events
$\om $ and the quantum state $ \Psi (\, \cdot \, | \vx) $ (the
class of equivalence $ \{\vp \}_{\vx} $) as a space $ \Om (\vx) $
of the elementary events. The observable $ \A $ is a random
variable

 $$ \vp\stackrel {\A} {\longrightarrow} A\equiv\vp(\A). $$

Let value of the observable be measured in experiment $\A\in\qqq $
and the device of the type $ \qqq $ be used. We denote the
measurable space of the elementary events by
$(\Om(\vx),\F_{\xi'})$. It corresponds to the quantum state
$\Psi(\,\cdot\,|\vx)$ and  to the $ \sss $-algebra $ \F_{\xi'} $
(to the measuring device of the type $ \qqq $). Let $P_{\xi'} $ be
a probabilistic measure on this space, i.e. $P_{ \xi'}(F) $ is
probability of the event $F\in\F_{\xi'}) $.

Let us consider that the event $ \aA $ is realized in experiment
if the registered value of the observable $ \A $ is no more $\aA$.
We denote probability of this event by $P_{\xi'} (\aA) =P (\vp:
\vp_{\xi'} (\A) \le\aA) $. Knowing probabilities $P_{ \xi'} (F) $,
we can find probability $P_{\xi'} (\aA) $ with the help of
corresponding summations and integrations. Distribution $P_{\xi'}
(\aA) $ is marginal for the probabilities $P_{\xi'}(F) $.

The observable $ \A $ can belong not only the subalgebra $ \qqq $
but also other maximal subalgebra $ \QQ_{\xi"} $. Therefore, for
definition of probability of event $ \aA $ we can use the device
of the type $ \QQ_{\xi"} $. In this case for probability we could
obtain other value $P_{\xi"} (\aA) $. However, experiment shows
that the probabilities do not depend on a used measuring device.
Therefore, we should accept one more postulate.

\

{\it Postulate 5:}

{\it Let the observable be $ \A\in\qqq\cap\QQ_{\xi"} $, then the
probability to find out the event $ \aA $ for the system which are
in the quantum state $ \Psi (\, \cdot \, | \vx) $, does not depend
on of the type used device, i.e. $P (\vp: \vp_{\xi'} (\A) \le\aA)
= P (\vp:\vp_{\xi"} (\A) \le\aA) $.}

\

Therefore, although the functional $ \vp $ can be multivalued, we
have the right to use notations $P (\vp:\vp (\A) \le\aA) $ for
probability of the event $ \aA $.

Let we have ensemble of the quantum systems which are in the
quantum state $ \Psi(\,\cdot\,| \vx) $. For this ensemble we carry
out a series of experiments in which the observable $ \A $ is
measured. We deal the finite set of the physical states in any
real series. In the ideal series this set can be denumerable. We
let $ \{\vp\}^A_{\vx} $ denote a random denumerable sample in the
space $ \Om(\vx) $ which contains all the physical states of the
real series. By the law of the large numbers (see for
example,~\cc{nev}) the probabilistic measure $P_{\A}$ in this
sample is uniquely determined by the probabilities
$P(\vp:\vp(\A)\le\aA) $.

The probabilistic measure $P_{\A} $ determines average value of
the observable $ \A $ in the sample $ \{\vp \}^A_{\vx} $:
 \beq {2}
\langle\A\rangle = \int_{\{\vp
\}^{A}_{\vx}}\,P_{\A}(d\vp)\,\vp(\A) \equiv \Psi (\A | \vx). \eeq
This average value does not depend on concrete sample, and is
completely determined by the quantum state $ \Psi(\,\cdot\,|\vx)$.

Formula~\rr{2} defines a functional (quantum average) on set
$\AAA_+$. We denote this functional also by $\Psi(\,\cdot\,|\vx)$.
The totality of all quantum experiments specifies that we must
accept the following postulate.

\

{\it Postulate 6:}

{\it The functional $ \Psi(\,\cdot\,|\vx)$ is linear on the set
$\AAA_+ $.}

\

\ \\ It implies that $$ \Psi(\A + \B|\vx) = \Psi (\A|\vx) +
\Psi(\B|\vx) \mbox { also in the case where } [\A, \B] \ne0. $$

Any element $ \R $ of the algebra \AAA {} is uniquely represented
as $ \R = \A+i\B $, where $ \A, \B\in\AAA _ + $. Therefore, the
functional $ \Psi (\, \cdot \, | \vx) $ can be uniquely extended
to a linear functional on \AAA: $ \Psi (\R | \vx) = \Psi (\A |
\vx) +i\Psi (\B | \vx) $.

Let us define norm of the element $ \R $ by equality $$\|\R \|^2 =
\sup_{\vx} \vx (\R^*\R). $$

Such definition is allowable. Due to the property (2.3) we have
$\| \R \|^2\ge 0 $. It follows from property (2.1) and Postulate 4
that if $ \|\R \|^2=0 $, then $ \R=0 $. Further, by virtue of
definition of the probabilistic measure
 $$ \Psi(\R^*\R | \vx) =
\int_{\{\vp \}^{R^*R}_{\vx}}\, P_{\R^*\R}(d\vp)\,\vp(\R^*\R)
\le\sup_{\vp_{\xi'}} \vp_{\xi'} (\R^*\R). $$
 If $ \R^*\R\in\QQQ$,
then $ \Psi (\R^*\R | \vx) = \vx (\R^*\R) $. Therefore,
 $$ \|\R\|^2 = \sup_{\vx}\vx (\R^*\R)=\sup_{\vx}\Psi(\R^*\R|\vx). $$
  Because $ \Psi (\, \cdot \, | \vx) $ is a linear positive
functional, the Cauchy-Bunyakovskii-Schwars inequality

$$ |\Psi(\R^*\s|\vx)\Psi(\s^*\R |\vx)|\le \Psi (\R^*\R | \vx) \Psi
(\s^*\s | \vx). $$ is satisfied.

Because $\vx ([\R^*\R]^2) = [\vx (\R^*\R)]^2 $, we have
$\|\R^*\R\| = \| \R \|^2 $. Therefore, if we complete the algebra
\AAA {} with respect to the norm $ \| \cdot \| $, then \AAA{}
turns out $C^* $-algebra~\cc{dix}.

Thus, a necessary condition of consistency of the postulate of the
linearity is the following strengthening of Postulate 1.

\

 {\it Postulate 1à:}

{\it The set of the dynamical variables is algebra which can be is
equipped with structure $C^* $-algebra.}

\

The reason that we did not accepted this formulation of the first
postulate initially because it follows from the experiment that
the observables have algebraic properties and the quantum mean
values have the linearity property. But mathematical relations
included in the definition of a $C^* $-algebra are not  directly
related to the experiment.

 \section {Time evolution and the ergodicity condition}

In the standard quantum mechanics the time evolution is determined
by the unitary automorphism
 \beq {4} \A (t) = \U^{-1}(t)\A(0)\U (t) \qquad \A (0) = \A, \eeq
where $ \A (t) $ and $ \U (t)$ are operators in some Hilbert
space. The operators of evolution $ \U (t) $ realize unitary
representation of one-parameter group. But for~\rr {4} to preserve
its physical meaning, it suffices to consider $ \A (t) $ and $ \U
(t) $ as elements of some algebra (in particular, of \AAA{}).

In our case, the evolution equation can be rewritten in terms of
physical states. We therefore accept

\

{\it Postulate 7}.

{\it A physical state of a quantum system evolves in time as
\beq{5} \vp (\A) \to \vp_t (\A) \equiv\vp (\A (t)), \eeq
 where $\A (t) $ is defined by Eq.~\rr{4}.}

\

Equation~\rr{5} describes time evolution of a physical state
entirely unambiguously. It is a different story, though, that an
observation allows determining the initial value $ \vp (\A) $ of a
functional only up to its belonging to a certain quantum state
$\{\vp \}_{\vx} $. Most of our predictions regarding the time
evolution of a quantum object are therefore probabilistic. In
addition, Eqs.~\rr{4} and~\rr{5} are valid only for systems that
are not exposed to first-class actions (in von Neumann's
terminology~\cc{von}), i.e., do not interact with a classical
measuring device.

We now return to the fifth and the sixth postulates. From the
experimental standpoint, these postulates are well justified. But
it is not quite clear whether they can be realized within the
mathematical scheme considered here. It turns out that these
postulates can be related to the time evolution of the quantum
system. For this, we must impose restrictions on the elements
$\U(t) $.

We now accept

\

 {\it Postulate 8}.

{\it The elements $ \U (t) $ are unitary elements, which have an
integral representation of the form
 \beq {014} \U (t) = \int \p(dE) \exp [iEt], \eeq
  where $ \p (dE) $ are orthogonal projectors.
The spectrum of $ \U (t) $ contains at least one discrete
nondegenerate value $E_0 $.}

\

Hereinafter integrations (and also limits) on algebra~\AAA{} are
understood in sense of the weak topology of
$C^*$-algebra~\cc{dix}.

Somewhat conventionally, we can represent $ \p (dE) $ as
 \beq {7}
\p(dE)=\p_p(dE)+\p_c(dE)=\sum_n\p_n\,\delta(E-E_n)\,dE+\p_c(dE.)
\eeq
 Here $ \p_p (dE) $ and $ \p_c (dE) $ concern to point and
continuous spectrums, accordingly. Besides, $ \p_n\p_m =
\p_m\p_n=0 $ for $m\ne n $, $ \p_n\p_c (dE) = \p_c (dE) \p_n=0 $.
The sum over $n $ in~\rr{14} must necessarily involve at least one
term ($n=0 $) with a nondegenerate value $E_0 $.

In addition to this last restriction, other requirements are
always assumed in considering any quantum mechanics model.
Requiring a discrete point in the spectrum does not seem too
restrictive either. For example, a one-particle quantum system can
have a purely continuous energy spectrum. But it can be considered
as a one-particle state of an extended system that can also be in
the vacuum state in addition to the one-particle state. The energy
spectrum of the extended system already has a discrete
nondegenerate point in the spectrum.

By the nondegeneracy of $E_0 $, we assume that the projector
$\p_0$ in decomposition~\rr {14} is one-dimensional. A projector
$\p $ is said to be onedimensional if it cannot be represented as
$$\p= \sum_{\al}\p_{\al}, \quad \p_{\al} \ne\p,\quad \p\p_{\al} =
\p _ {\al} \p = \p _ {\al}. $$

Let us remark that, if two elements $ \A_1 $ and $ \A_2 $ of
algebra~\AAA{} have identical spectral representation, then they
obey the fourth postulate. Therefore, such elements coincide.

We call physical state $ \vp_0^{\al} $ a ground state if
$\vp_0^{\al} (\p_0) =1 $.

\

{\sc Statement.} {\it If $ \A\in\AAA_+ $, then
$\A_0\equiv\p_0\A\p_0 $ has the form $ \A_0 = \p_0 \, \po (\A) $,
where $ \po (\A) $ is the linear, positive functional. It
satisfies the normalization condition $ \po (\I) =1 $.}

\

{\sc Proof.} Because $ [\A_0, \p_0] =0 $ it follows that $ \A_0 $
and $ \p_0 $ have the common spectral decjmposition of unity.
Since the projector $ \p_0 $ is one-dimensional, the spectral
decomposition $ \A_0 $ has the form $ \A_0 = \p_0 \, \po (\A) +
\A'_0 $ where $ \A'_0 $ is orthogonal to $ \p_0 $. Therefore,
$\A_0 = \p_0\A_0 = \p_0\p_0\po (\A) + \p_0\A'_0 = \p_0\po (\A) $.

Linearity: $$ \p_0\po (\A + \B) = \p_0 (\A + \B) \p_0 = \p_0 \,
\po (\A) + \p_0\po (\B.) $$ From here follows $ \po (\A + \B) =
\po (\A) + \po (\B) $.

By linearity, the functional $ \po (\A) $ can be expanded to the
algebra \AAA, $ \po (\A+i \, \B) = \po (\A) +i\po (\B) $, where
$\A, \B\in\AAA_+ $.

Positivity: $$\po(\R^*\R) = \vp_0^{\al} (\p_0\po (\R^*\R)) =
\vp_0^{\al} (\p_0\R^*\R\p_0) \geq 0, $$ by virtue of the property
(2.3).

Normalization: $$\po (\I) = \vp_0^{\al} (\p_0\po (\I)) =
\vp_0^{\al} (\p_0\I\p_0) =1 $$.

To find the physical meaning of the functional $ \po $, it is
necessary to consider an element $ \Aa $ in the algebra \AAA {}
that corresponds to an observable $ \A $ averaged in time.
\beq{8}
\Aa = \lim _ {L\to\infty} \frac {1} {2L} \int^L _ {-L} dt \, \A
(t) = \lim _ {L\to\infty} \frac {1} {2L} \int^L _ {-L} dt \, \U ^
{-1} (t) \, \A \, \U (t.) \eeq

It is possible to show (see~\cc {slav4}) that $$\po (\A) =
\vp_0^{\al} (\Aa). $$ That is, the value of the observable $ \A $
in the quantum ground state $ \po $ is equal to the value of the
observable $ \Aa $ in the physical ground state $ \vp_0^{\al} $.
This value is the same in all physical ground states.

The functional $ \po $ has all the properties that must be
possessed by a functional determining quantum mean values. It is
linear, is positive, and is equal to unit on the unit element. In
addition it is continuous, as a linear functional on the
$C^*$-algebra. Therefore, we can accept the ergodicity axiom.

\

{\it Postulate 9:}

{\it The mean value of an observable $ \A $ in the quantum ground
state is equal to the value of the observable $ \Aa $ (the
observable $ \A $ average in time) in any physical ground state.}

\

Thus, averaging in quantum ensemble can be reduced to averaging in
time. We note that the Postulate 9 does Postulates 5 and 6
superfluous.

To construct the standard mathematical formalism of quantum
mechanics, we can now use the canonical construction of
Gelfand-Naimark-Segal (GNS) (see, e.g.,~\cc{emch}).

We consider two elements $ \R, \, \s\in\AAA $ equivalent if the
condition $ \po\left (\K^* (\R-\s) \right) =0 $ is valid for any
$\K\in\AAA $. We let $ \Phi (\R) $ denote the equivalence class of
the element $ \R $ and consider the set $ \AAA (\po) $ of all
equivalence classes in \AAA. We make $ \AAA (\po) $ a linear space
setting $a\Phi (\R) +b\Phi (\s) = \Phi (a\R+b\s) $. The scalar
product in $ \AAA (\po) $ is defined as $ \left (\Phi (\R),
\Phi(\s) \right) = \po (\R^*\s) $. This scalar product generates
the norm $ \| \Phi (\R)\|^2 = \po(\R^*\R) $ in $ \AAA (\po) $.

Completion with respect to this norm makes $ \AAA (\po) $ a
Hilbert space. Each element $ \s $ of the algebra \AAA {} is
uniquely assigned a linear operator $ \Pi _ {\Psi} (\s) $ acting
in this space as $ \Pi _ {\Psi} (\s) \Phi (\R) = \Phi (\s\R) $.

\section {Examples}

To illustrate the above, we  consider two simple examples.

First we consider a quantum system whose observable quantities are
described by Hermitian $2\times 2 $ matrices. The Hamiltonian $
\HH $ and the elements $ \p_0 $, $ \A $ are given by $$\HH =\lt
[\begin {array} {cc} E_0 & 0 \\ 0 &-E_0 \end{array} \rt], \qquad
\p_0 =\lt [\begin {array} {cc} 0 & 0 \\ 0 & 1
\end{array} \rt], \qquad \A =\lt [\begin {array} {cc} a & b \\ c
& d\end{array} \rt]. $$
 Obviously, $ \p_0 \,\po (\A) = \p_0 \,\A\,\p_0 =\p_0 \, d $, i.e.,
  \beq {13} \po (\A) =d. \eeq
   In addition,
\beq {14} \Aa =\lim_{L\to\infty} \frac{1}{2L} \int^L_{-L}dt\,e^{-
iE_0t\tau_3} \, \A \, e ^ {iE_0t\tau_3} = \lt[\begin {array} {cc}
a & 0 \\ 0 & d \end{array} \rt], \eeq
 where $ \tau_i $ are Pauli matrices.

All physical states can easily be constructed. We consider a
Hermitian matrix $ \A $, i.e., with $a^* = a $, $d^* = d $, and
$c=b^* $. Any such matrix can be represented as
 \beq {15}
\A=r_0\I+r \,\hat\tau (\bar n), \eeq where $\bar n$ is the unit
three-dimensional vector, $\hat\tau(\bar n)=(\bar\tau \,\bar n) $.
For Eq.\rr{15} to be valid, we must
 $$
r =\lt (\frac{(ad)^2}{4} +b\, b^*\rt)^{1/2}, \quad r_0
=\frac{a+d}{2}, $$ $$ n_1 =\frac {b+b ^ *} {2r}, \quad n_2
=\frac{b-b^*} {2ir}, \quad n_3 =\frac {a-d} {2r}. $$
 The commutator of the matrices $\hat\tau(\bar n),\hat\tau(\bar n') $
is nonvanishing for $ \bar n '\ne\pm\bar n $. Therefore, each
matrix $ \hat\tau (\bar n) $ (up to a sign) is a generator of a
real maximal commutative subalgebra. Because $ \hat\tau (\bar n)
\hat\tau (\bar n) = \I $, the spectrum of $ \hat\tau (\bar n) $
consists of two points $ \pm1 $.

Let $ \{f (\bar n) \} $ be the set of all functions taking the
values $ \pm1 $ and such that $f (-\bar n) =-f (\bar n) $. A
physical state is described by a functional whose value coincides
with one of the points in the spectrum of the corresponding
algebra element. For each point of the spectrum, there exists an
appropriate functional. Therefore, to the set of physical states,
there corresponds a set of functionals defined by
 $$ \vp\left(\hat\tau (\bar {n}) \right) =f (\bar {n}). $$
Taking properties \rr{1} into account (which must be possessed by
each physical state), we obtain
 \beq {16} \vp (\A) =r_0+r \, f(\bar {n}). \eeq
  The ground state is any functional $ \vp_{0\al} $
such that $$ f (n_1=0, n_2=0, n_3=1) =-1. $$ Substituting the
element $ \Aa $ (\rr {14}) in \rr {16}, we obtain
 $$
\vp_{0\al}(\Aa) = \frac {a+d} {2} - \frac {a-d} {2} =d. $$
 This agrees with \rr{13}.

Because all maximal commutative subalgebras have one independent
generator in this model, the physical states are described by the
single-valued functionals. If there were several generators, then
multivalued functionals would inevitably arise. It corresponds to
the result received (in other terms) by Kochen and Specker
\cc{ksp}.

As the second example we consider a harmonic oscillator.

In this case the algebra of dynamical variables is algebra with
two noncommuting Hermitian generators $ \Q $ and $ \PP $
satisfying the commutative relation $$ [\Q, \PP] =i. $$  Time
evolution in the algebra controls the Hamiltonian $$\HH=1/2 (\PP^2
+\nu^2\Q^2). $$ The elements $ \Q $, $ \PP $ and $ \HH $ are
unbounded. Therefore, they do not belong to the $C^* $-algebra.
However, their spectral projectors are elements of the
$C^*$-algebra, i.e. $ \Q $, $ \PP $ and $ \HH $   is the elements
joined to  the $C^* $-algebra . Thus, in this case the
\AAA-algebra  is a $C^* $-algebra with the joined elements.

It is convenient to turn from the Hermitian elements $ \Q $ and $
\PP $ to elements
$$\am=\frac{1}{\sqrt{2\nu}}(\nu\Q+i\PP), \qquad
\ap=\frac{1}{\sqrt{2\nu}}(\nu\Q-i\PP)$$
 with the commutative relation
 \beq {17} [\am, \ap] =1 \eeq
  and simple time dependence
   $$\am (t) = \am\exp (-i\nu t), \qquad \ap(t) = \ap\exp (+i\nu t). $$

Let us calculate a generating functional for Green functions. In
standard quantum mechanics the $ n $-time Green function is
defined by the equation
 \beq {18}
  G (t_1,\dots t_n) = \langle 0|T(\Q (t_1) \dots \Q (t_n))|0\rangle, \eeq
where $T $ is a operator of chronological ordering and $|0\rangle$
is a quantum ground state.

According to  Postulate 9 in the proposed approach the Green
function is defined by the equation
 \beq {19}
  \p_0 T (\Q (t_1)\dots \Q (t_n)) \p_0=G (t_1, \dots t_n) \p_0, \eeq
where $ \p_0 $ is a spectral projector $ \HH $ corresponding to
the minimal value of energy.

It is easy to make sure that $ \p_0 $ can be represented in form
\beq {20} \p_0 = \lim _ {r\to \infty} \exp (-r\ap\am). \eeq
 As well as earlier, here the limit is understood in sense of weak
topology of the $C^* $-algebra.

 First we  prove the auxiliary statement:
  \beq {21} \J = \lim _ {r_1, r_2\to \infty} \exp (-r_1\ap\am)
(\ap) ^k (\am) ^l\exp (-r_2\ap\am) =0. \eeq
 Let $ \Psi $ be an any positive linear functional. Then
  \beq {22}
\Psi (\J) = \lim_{r_1, r_2\to \infty} \exp (-r_1k-r_2l) \Psi
((\ap) ^k\exp (-r_1\ap\am) \exp (r_2\ap\am) (\am) ^l). \eeq
 Here, we have used a continuity of the functional $ \Psi $ and the
commutative relation~\rr{17}. Further,
 \bea{23} |\Psi(\J)|&\leq&
\lim_{r_1,r_2\to \infty}\exp(-r_1k-r_2l) |\Psi(
(\ap)^k\exp(-r_1\ap\am)\exp(-r_1\ap\am)(\am)^k)|^{1/2}\nn
&\times&\quad
|\Psi((\ap)^l\exp(-r_2\ap\am)\exp(-r_2\ap\am)(\am)^l)|^{1/2} \nn
&\leq& \lim_{r_1,r_2\to \infty}\exp(-r_1k-r_2l)
|\Psi((\ap)^k(\am)^k)|^{1/2} |\Psi((\ap)^l(\am)^l)|^{1/2} \eea
 Here we considered that $\|\exp(-r\ap\am)\|\le 1$.  It  follows
from~\rr{23}  that  $ | \Psi (\J) | =0 $, i.e. it is
valid~\rr{21}.

We now  prove~\rr{20}. In terms of the elements $ \ap $, $ \am $
the Hamiltonian $ \HH $ has form $ \HH =\nu (\ap\am+1/2) $.
According~\rr{21},
 $$ \lim _ {r_1, r_2\to \infty} \exp (-
r_1\ap\am) \HH\exp (-r_2\ap\am) = \frac {\nu} {2} \lim _ {r_1,
r_2\to \infty} \exp (-(r_1+r_2) \ap\am) = \frac {\nu} {2} \lim _
{r\to \infty} \exp (-(r) \ap\am). $$
 It proves the equality~\rr{20}.

It follows from the equation~\rr{19} that
 \beq {24}
G (t_1, \dots t_n) \p_0 =\lt.\lt (\frac {1} {i} \rt) ^n\frac
{\delta^n} {\delta j (t_1) \dots\delta j (t_n)} \p_0 T\exp\lt
(i\int ^\infty _ {-\infty} dt \, j (t) \Q (t) \rt) \p_0\rt | _
{j=0}. \eeq

 By the Wick theorem (see~\cc{bog})
    \bea{25}
&T\exp\lt( i\int^\infty_{-\infty} dt\,j(t)\Q(t)\rt)=\nn& =
\exp\lt( \frac{1}{2i}\int^\infty_{-\infty} dt_1dt_2\,
\frac{\delta}{\delta\Q(t_1)}D^c(t_1-t_2)\frac{\delta}{\delta\Q(t_2)}\rt)
:\exp\lt( i\int^\infty_{-\infty} dt\,j(t)\Q(t)\rt):.
 \eea
 Here $: \;: $ is operation of normal ordering and
  $$D^c (t_1-t_2) = \frac {1} {2\pi} \int dE \,\exp\lt (-i (t_1-t_2)
E\rt) \frac {1} {\nu^2-E^2-i0} $$

Carrying out a variation over $ \Q $ in the right-hand
side~\rr{25} and taking into account~\rr{21}, we have
  \begin{eqnarray*}
\p_0T\exp\lt( i\int^\infty_{-\infty} dt\,j(t)\Q(t)\rt)\p_0&=&
\exp\lt(- \frac{1}{2i}\int^\infty_{-\infty} dt_1dt_2\,
j(t_1)D^c(t_1-t_2)j(t_2)\rt)\nn \times \p_0:\exp\lt(
i\int^\infty_{-\infty} dt\,j(t)\Q(t)\rt):\p_0 &=& \p_0  \exp\lt(-
\frac{1}{2i}\int^\infty_{-\infty} dt_1dt_2\,
j(t_1)D^c(t_1-t_2)j(t_2)\rt).
 \end{eqnarray*}

  Comparing with~\rr{24}, we obtain
  $$ G (t_1\dots t_n) = \lt.\lt (\frac {1} {i} \rt) ^n\frac {\delta^n Z
(j)} {\delta j (t_1) \dots \delta j (t_n)} \rt | _ {j=0}, $$
 where
$$ Z (j) = \exp\lt (\frac {i} {2} \int ^\infty _ {-\infty}
dt_1dt_2 \, j (t_1) D^c (t_1-t_2) j (t_2) \rt) $$ is the
generating functional.

This method of calculation of a generating functional for Green
functions is easily to generalize for more substantial quantum
models, in particular, quantum-field models.

 \section {Bell inequality}

We now investigate how the measurability condition on the
probability space is manifested in the important case of the
derivation of the Bell inequality~\cc{bell}. There are many forms
of this inequality. Hereinafter, we refer to the version proposed
in~\cc{chsh}. This variant is usually designated by the
abbreviation CHSH.

Let particle with the spin 0 decay into two particles $A $ and $B
$ with spin 1/2. These particles move apart, and the distance
between them becomes large. The projections of their spins are
measured by two independent devices $D_a $ and $D_b $. Let the
device $D_a $ measures the spin projection of the particle $A $ on
$a $ direction, and the device $D_b $ measures the spin projection
of the particle $B $ on the $b $ direction. We let $ \A $ and $ \B
$ denote the corresponding observables and let $A_a $ and $B_b $
denote the measurement results.

Let us assume that the initial particle has some physical reality
that can be marked by the parameter $ \lll $. We use the same
parameter to describe the physical realities for the decay
products. Accordingly, it is possible to consider measurement
results of the observables $ \A $, $ \B $ as the function $A_a
(\lll) $, $B_b (\lll) $ of the parameter $ \lll $. Let the
distribution of the events with respect to the parameter $ \lll $
be characterized by the probabilistic measure $P (\lll) $:
 $$ \int dP (\lll) =1, \qquad 0\leq P (\lll) \leq 1. $$

We introduce the correlation function $E (a, b) $:
 \beq {29} E (a,b) = \int dP (\lll) \, A_a (\lll) \, B_b (\lll) \eeq
 and consider the combination
  \bea {30}I&=&|E (a, b) -E (a, b ') | + |E (a ', b) +E (a ', b ') | = \nn{}
 &=& \lt |\int P(d\lll)\,
A_a(\lll)\,[B_b(\lll)-B_{b'}(\lll)]\rt|+ \lt |\int P(d\lll)\,
A_{a'}(\lll)\, [B_b(\lll) +B_{b'}(\lll)]\rt|.\eea
 The equalities
  \beq {31} A_a (\lll) = \pm1/2, \quad B_b(\lll) = \pm1/2 \eeq
  are satisfied for any directions $a $ and $b$. Therefore,
   \bea{32} I & \le &\int
P(d\lll)\,[|A_a(\lll)|\,|B_b(\lll)-B_{b'}(\lll)|+ |A_{a '} (\lll)
| \, |B_b (\lll) +B_{b '} (\lll) |] =\nn{}
 &=&1/2 \int P(d\lll) \,
[|B_b(\lll)-B_{b'}(\lll)|+|B_b(\lll)+B_{b'}(\lll)|].\eea

Due to the equality~\rr{31} for each $ \lll $ one of the
expressions \beq{33}
 |B_b(\lll) -B_{b'} (\lll) |, \qquad |B_b (\lll) + B_{b'}
(\lll) | \eeq
 is equal to zero and the other is equal to unity.
Here it is crucial that the same value of the parameter $ \lll $
appears in both expressions. Hence, the Bell inequality (CHSH)
then follows:
  \beq {34} I \leq 1/2\int dP (\lll) =1/2. \eeq

The correlation function can be easily calculated within standard
quantum mechanics. We obtain
 $$ E (a, b) =-1/4\cos\theta _ {ab},$$
where $ \theta _ {ab} $ is the angle between the directions $a$
and $b $. For the directions $a=0 $, $b = \pi/8 $, $a' = \pi/4 $,
$b' =3\pi/8 $ we have $$ I=1/\sqrt {2}. $$ It contradicts the
inequality~\rr{34}.

Experiments that have been performed corresponded to
quantum-mechanical calculations and did not confirm the Bell
inequality. These results have been interpreted as decisive
argument against the hypothesis of the existence of local
objective reality in quantum physics. It is easy to see that if
the probability theory is properly applied to quantum system, then
the above derivation of the Bell inequality is invalid.

Because the $ \sss $-algebra and accordingly probability measure
depend on the measuring device used in a quantum case, it is
necessary to make replacement $dP (\lll) \to dP _ {\A\B}(\vp) $ in
the equation~\rr{29}. If we are interested in correlation function
$E (a', b') $,  it is necessary to make replacement $P(d\lll) \to
P_{\A'\B'} (d\vp) $ in the equation~\rr{29}. Although we used the
same symbols $d\vp $ in both cases for notation of the elementary
volume in the space of the physical states, it is necessary to
remember  that sets of the physical states corresponding $d\vp $,
are different. The matter is that these sets should be elements of
the $ \sss $-algebras. If observables $ \A $, $ \B $ are
incompatible with observables $\A'$, $ \B '$, then $\sss
$-algebras are different. Moreover, there are no physically
allowable $ \sss $-algebra which has these algebras as
subalgebras.

Besides in experiment we deal not with the complete probability
space $ \Om (\vx) $ but with random denumerable sample
$\{\vp\}^{AB}_{\vx} $. Finally, the equation~\rr{29} should be
replaced on
$$E(a,b)=\int_{\{\vp\}^{AB}_{\vx}}P_{\A\B}(d\vp)\vp(\A\B). $$

Accordingly, the equation~\rr{30} now has the form
 \begin{eqnarray*}
I&=& \lt |\int_{\{\vp\}^{AB}_{\vx}}P_{\A\B}(d\vp)\, \vp(\A\B)-
\int_{\{\vp\}^{AB'}_{\vx}} P_{\A\B'}(d\vp)\,\vp(\A\B')\rt|+ \nn{}
& +& \lt |\int_{\{\vp\}^{A'B}_{\vx}} P_{\A'\B}(d\vp)\, \vp(\A'\B)
+ \int_{\{\vp\}^{A'B'}_{\vx}} P_{\A'\B'}(d\vp)\, \vp(\A'\B')\rt|.
  \end{eqnarray*}

If the directions $a $ and $a' $ ($b $ and $b' $) are not parallel
to each other, then the observables $ \A\B $, $ \A\B' $, $\A'\B$,
$ \A'\B' $ are mutually incompatible. Therefore, there is no
physically acceptable universal $ \sss $-algebra that corresponds
to the measurement all these observables. It follows that there is
no probability measure common for these observables. Besides, the
sets $ \{\vp \}^{AB}_{\vx}$, $ \{\vp \}^{AB'}_{\vx} $,
$\{\vp\}^{A'B}_{\vx} $, $ \{\vp \}^{A'B'}_{\vx} $ are different
random denumerable samples from continual space $ \Om (\vx) $. The
probability of their  crossings is equal to zero. Therefore, the
probability of occurrence of combinations of the type~\rr {33} is
equal to zero. As a result, the reasoning which have led to to an
inequality~\rr{34},  appears unfair for the physical states.

Thus, the hypothesis that local objective reality does exist in
the quantum case does not lead to the Bell inequalities.
Therefore, the numerous experimental verifications of the Bell
inequalities that have been undertaken in the past and at present
largely lose theoretical grounds..

\section {The possible carrier of "the objective local reality"}

In the previous sections of the paper we tried to show that,
contrary to a popular belief, the mathematical formalism of the
standard quantum mechanics does not contradict the hypothesis on
existence of an objective local reality. In the developed approach
this reality is identified with concept "the physical state". The
mathematical essence of this concept is defined quite uniquely
(the functional $ \vp $). It would be desirable to have some
physical filling of this concept.

In a present section we heuristically consider one of the
variants~\cc{slav5} of such filling. This variant is not unique.
At the same time, it gives simple and obvious interpretation to
the problem phenomenon such as a collapse of the quantum state.

Within the framework of the algebraic approach the quantum object
is a finite region in the space-time, which the certain
noncommutative local algebra (algebra of quantum local
observables) is associated with.

On the other hand, the quantum object is a source of some field.
Obviously, any quantum object is a source of the gravitational
field. Till now all attempts consistently to quantize the
gravitational field were not crowned with success. Probably, it is
related to that the gravitational field is classical, i.e. the
corresponding algebra of observables is commutative.

Probably, quantum objects are also sources of other classical
fields, in particular, the classical electromagnetic field. This
field should be very weak. In this case for microobjects it is
unobservable on background of the quantum electromagnetic field.

It is natural to assume that the classical field radiated by a
microobject, is coherent to this object. Thus, it is a carrier of
the information about the object. If the microobject is
multipartial then the radiated field is coherent to both the
separate parts of the object, and to all collective. In this case
the field is a carrier of the information about correlations.

Macroscopic bodies can act on the radiated classical field
essentially. This action can be two types. The first type is
action which destroys the coherence of the field. Such action is
irreversible. The second type is action which preserves the
coherence. So the mirror acts on radiation. Usually, both type of
interactions are present at measuring devices. The coherence is
preserved in the analyzer and destroyed in the detector.

On account of weaknesses the classical field exercises negligibly
small effect on both micro- and macroobjects. Unique exception is
the microobject, coherent to this field. In this case the action
is resonant. Therefore, even very small action can lead to
appreciable result if the quantum object is in the state of the
bifurcation. That is, in the state where without taking into
account this action several variants of the further evolution are
possible. In this case such action can play a role of random force
which forces the quantum object to choose one of the variants. In
this sense the classical field can be considered as a pilot field.
It is possible to assume that the configuration of the classical
field, coherent to the quantum object, is the objective reality
which determines the physical state of this object.

For an illustration of this phenomenon we discuss experiment whose
scheme is represented on Figure 1.
 \begin {center} \begin {picture} (120,70) \put (25,10) {\vector (1,0) {30}}
\put (25,40) {\vector (1,0) {30}} \put (55,10) {\vector (1,0)
{55}} \put (25,55) {\vector (0,-1) {30}} \put (25,25) {\line
(0,-1) {15}} \put (85,40) {\vector (0,-1) {15}} \put (25,65)
{\vector (0,-1) {10}} \put (85,25) {\vector (0,-1) {23}} \put
(90,10) {\line (1,0) {5}} \put (55,40) {\line (1,0) {30}} \put
(25,55) {\oval (5,10)} \put (55,10) {\oval (10,5)} \put (100,10)
{\oval (10,5)} \put (25,40) {\circle {8}} \put (85,10) {\circle
{8}} \put (30,35) {$1 $} \put (80,35) {$2 $} \put (30,15) {$3 $}
\put (80,15) {$4 $} \put (30,55) {$D_1 $} \put (55,15) {$D_2 $}
\put (100,15) {$D_3 $} \thicklines \put (22.1,13) {\line (1,-1)
{5.8}} \put (22.1,43) {\line (1,-1) {5.8}} \put (82.1,13) {\line
(1,-1) {5.8}} \put (82.1,43) {\line (1,-1) {5.8}} \end {picture}

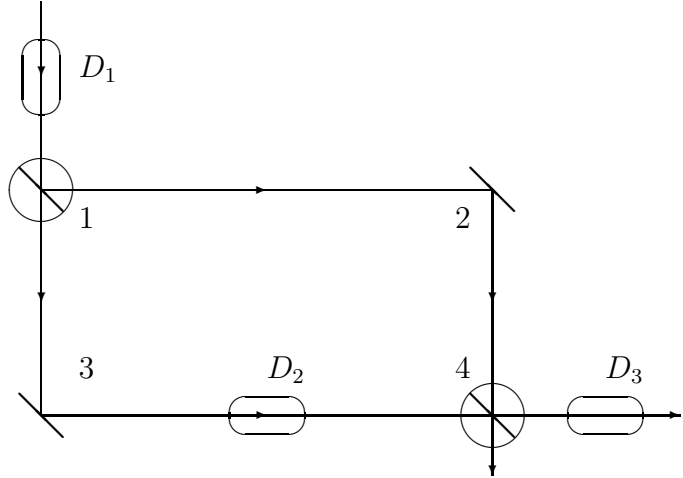
Figure 1. \end {center}

The device consists of four mirrors (1,2,3,4) and three detectors
($D_{1}, D_{2}, D_{3} $). The mirrors 1 and 4 are semipermeable.
The detectors $D_{1} $ and $D_{3} $ are necessary only for
registration of photons. The detector $D_{2} $ plays central role
in the phenomenon of collapse. At the device the photon and the
coherent classical field either are reflected from mirrors, or
pass through them. After reflection from the mirrors the phase
changes by $ \pi/2 $, at passage through a semipermeable mirror
the phase does not change.

Second, the mirror 1 is a point of the bifurcation for the photon.
Without taking into account interaction with the oscillations
excited by the classical field in the mirror, both channels for
the photon are equally allowable. Oscillations are very weak but
they are coherent to the photon. Therefore, interaction is
resonant. Due to this interaction the information which is stored
in the classical field (the physical state) dictates to the photon
a choice of the channel.

From the point of view of quantum mechanics this choice is random.
The fact is that quantum mechanics deals not with physical state,
but only with its generalized characteristic --- quantum state.
The various configurations of the coherent classical field
correspond to the same quantum state. On the other hand, with the
help of preliminary measurements we can receive an information
only about the quantum state. Therefore, the choice of the routes
by the photon is random for us.

The phases of the photon and separate parts of the classical field
can vary when they pass channels, but their coherency is kept.
According to rules of the classical optics in the mirror 4 the
separate parts of the classical field interfere so that after the
mirrors 4 the field does not propagate to the detector $D_3 $.
Physically the classical field raises the small collective
oscillations in the mirror 4 coherent to the field. The scattering
occurs on these oscillations. The photon also is coherent to these
oscillations and scatters the same as the classical field.
Therefore, it also does not hit the detector $D_3 $.

We now consider the second variant of the experiment when the
detector $D_{2} $ is switched on. In the mirror 1 everything
happen the same way as in the first variant. Two scenarios are
further possible, in which the photon goes by the route 1-3-4 or
does by the route 1-2-4.

In the scenarios with the route 1-3-4 the photon hits the detector
$D_{2} $. There the photon participates in interaction with the
classical device. The device goes out of unstable equilibrium due
to interaction with the photon. The catastrophic process develops
in the device. This process has macroscopically observable result
and the quantum object is registered.

The detector exerts strong action on the photon. Its state
changes, and it loses a coherency with earlier radiated classical
field. Again radiated classical field is coherent to the photon in
the new state. At the same time, the field in the channel 1-2-4 is
not coherent to the photon. Because only the coherent classical
field determines the physical state of the quantum object, the
field in the channel 1-2-4 is effectively lost for the photon.

This process results in a sharp modification of the coherent
classical field of the quantum object. The quantum state of the
object also changes sharply thereof. This phenomenon has all
features of the collapse. However, any inconsistency with a
relativity theory does not arise, as in the channel 1-2-4 the
classical field does not change. The modifications happen in the
channel 1-3-4. Thus, the classical fields in the channels 1-2-4
and 1-3-4 do not disappear in the collapse, but these fields lose
the coherence with each other. Therefore, in the mirror 4
interference is absent. Latter corollary agrees with the corollary
adduced in the review~\cc{nam}.

Let us consider now the second scenario in which the photon goes
by the route 1-2-4, and the photon-free classical field goes
through the detector $D_2 $. This field exercises negligibly small
effect on the detector. In this case the cause generating
catastrophic process in the detector is absent. Any macroscopicaly
observable of the reaction of the device is not present.

On the contrary, the action of the detector affects strongly the
classical field in the channel 1-3-4. This field loses coherence
with the field and the photon in the channel 1-2-4. The situation
is the same as in the first scenario.

Quite similarly it is possible to interpret so-called
delayed-choice experiment~\cc{wheel}. It is usually considered
that this experiment testifies to absence of the local physical
reality in the quantum phenomena.

We now can look at the experiment double-slits~\cc{ton} in a new
fashion. The distinct interference pattern is observed in this
experiment. If to reject any verbal ornaments standard
interpretation of this experiment is reduced to the following. Up
to the slits the {\it indivisible} quantum object passes
simultaneously through the slits divided by the macroscopic
distance then it again becomes {\it indivisible}.

The proposed approach allows to interpret this experiment much
more evidently. The indivisible quantum object hits  one of the
slits and scatters on it. It can scatter in region behind the slit
or in the opposite side. From the point of view of the standard
quantum mechanics this process is random. In terms of the present
paper it means that the slit is region of the bifurcation for the
quantum object. In this region the behaviour of the concrete
quantum object is determined by the random force, i.e. by
classical, coherent to the quantum object, the field in the area
of the slit.

According to rules of classical physics the part of the classical
field, scattered on both slits, interfere among themselves.
Therefore, presence of the second slit influences structure of the
field in the area of the first slit. Accordingly, presence of the
second slit influences on chance of passage of the quantum object
through the first slit. Thus, the ensemble of the quantum objects,
which have passed through simultaneously open slits, is not a mix
of two ensembles of the quantum objects, which have passed through
any one of the slits when another is closed. As a result there is
the interference pattern. If we place the detector in one of
slits, it disturbs the coherency of the parts of the classical
field, scattered on the different slits, and the interference
becomes impossible.

It is possible to look at this experiment from slightly other
position. It is possible to use these two slits as the device
determining localization of the quantum object. In this case we
can consider that if the quantum object is found out behind the
plane of the slits at the moment of passage of these slits, then
it is located inside one or two slits depending on what slits are
open. Permeability of each slit depends on that, other slit is
open or closed. Therefore, corollary about localization of the
concrete quantum object depends not only on properties of the
object but also on properties of the measuring device. Whether
both slits or only one of them are open in a present concrete
case. It is possible such, paradoxical from the point of view of
classical physics situation when two open slits appear
impenetrable for the quantum object which could pass through one
open slit when another closed.

 \section {Conclusions}

In principle, a physical state can be considered as a special
hidden parameter~\cc{bohm}. Hidden parameters have had a "bad
reputation" in quantum mechanics since the works of von Neumann.
In his illustrious monograph~\cc{von}, von Neumann argued that
models with hidden parameters are incompatible with the
fundamentals of quantum mechanics. As basic fundamentals of
quantum mechanics, he accepted, in particular, axioms: Linear
operators in the Hilbert space correspond to quantum observables;
the state of quantum system is described by a linear functional
(density matrix).

In the approach proposed here these statements are considerably
weakened. It is supposed only that observables are elements of
some algebra. They become linear operators only in representation
which is generated by linear functional of a special form (quantum
average). The physical state of the quantum system is described by
the nonlinear functional. The von Neumann's reasonings are not
valid for this functional.

Essentially, it was shown in~\cc{von} that the linearity of the
state is incompatible with causality and the hypothesis on hidden
parameters. Therefore, von Neumann concluded that there is no
causality at the microscopic level while it appears at the
macroscopic level because of averaging over a large number of
noncausal events.

The approach formulated in this work allows avoiding this conflict
in just the opposite way. It can be said that causality does exist
at the level of individual microscopic phenomena, but linearity
does not. The linearity of the (quantum) state appears because of
averaging over the quantum ensemble. The passage from individual
phenomena to quantum ensembles implies the passage from the
initial determinism to the probabilistic interpretation.

Because of the multivaluedness of the functional $ \vp $, the
conditions of the Kochen and Specker no-go theorem are not
satisfied for this hidden parameter. Finally, the conditions of
Bell's theorem are not satisfied for the functional $ \vp $.
Therefore, the arguments that are usually adduced by opponents of
the use of hidden parameters in quantum mechanics become
inapplicable for the physical state.

Regarding the multivaluedness of the functional $ \vp $, the
physical process that is usually called a measurement should
rather be called an observation. The term "observation" better
expresses the fact that the readout  of the device  has two
causes: the physical state of the observed object, and the type of
the device.

It then follows that the concept of a local objective reality and,
traced back to Bohr, "the situational (contextual) approach" are
not in such antagonistic contradiction as it is considered to be.
It is quite allowable that there is a physical reality which is
inherent to the quantum object under consideration and which
predetermines the result of any experiment. However, this result
can depend on conditions in which this experiment is carried out.
One of these conditions is the classical characteristic (type) of
the measuring device, which is used in a concrete case.

The quantum logic is not used in the present paper. Actually, the
quantum logic is a formulary for subspaces of the Hilbert space
and for projectors on these subspaces. By the way, these rules are
obtained with the help of the formal classical logic. They are
useful to the description of the Hilbert space. However it cannot
be the basis for attributing to these rules of general character
in the question "what is a true". Namely this question is in the
competence of  logic.

The Hilbert space is not a primary element in the proposed
approach to quantum mechanics. It appears as some by-product
alongside with others. Therefore, "quantum logic" is not the tool
which is useful in the present approach.

 \begin {thebibliography} {99}

\bb {von} Von~Neumann J.  Mathematical Foundation of Quantum, New
York: Mechanic Prentice-Hall  (1952).  \vspace {-2.5mm}

\bb{men} M.B.~Mensky. Physic-Uspekhi, 43 (2000) 585.
\vspace{-2.5mm}

\bb{ever} H.~Everett. Rev.Mod.Phys., 29 (1957) 454.
\vspace{-2.5mm}

\bb {emch}G.G.~Emch.  Algebraic Methods in Statistical Mechanics
and Quantum Field Theory, Wiley-Interscience, a Division of John
Wiley  and Sons, INC, New York (1972). \vspace {-2.5mm}

\bb{hor} S.S.~Horuzhy. Introduction to Algebraic Quantum Field
Theory, Kluwer, Dordrecht (1990).\vspace{-2.5mm}

\bb{blot}N.N.~Bogoliubov, A.A.~Logunov, A.I.~Oksak, and
I.T.~Todorov. General Principles of Quantum Field Theory, Kluwer,
Dordrecht (1990). \vspace{-2.5mm}

\bb{rud}W.~Rudin.   Functional Analysis, New York: McGraw-Hill
Company, (1973) \vspace{-2.5mm}

\bb{zeil}A.~Zeilinger.   Rev. Modern. Phys.,  71 (1999)  S289.
\vspace {-2.5mm}

\bb{jord}P.~Jorrdan.  Comm. Math. Phys.,  80  (1933) 285.
\vspace{-2.5mm}

\bb{slav1} D.A.~Slavnov.   Theor. Math. Phys., 129 (2001) 87;
D.A.~Slavnov  ArXiv: quant-ph/0101139.\vspace {-2.5mm}

\bb{slav2}D.A.~Slavnov.  Theor. Math. Phys., 132 (2002) 1264;
\vspace {-2.5mm}

\bb{ksp}  S.~Kochen,  E.P.~Specker.   Journ. of Mathematics and
Mechanics,   17   (1967) 59. \vspace {-2.5mm}

\bb{hom}  D.~Home,  M.A.B.~Whitaker.  Phys. Rep.,   210 (1992)
223. \vspace{-2.5mm}

\bb{kol}  A.N.~Kolmogorov,  Foundation of the theory of
probability, Chelsea, New York (1956).\vspace{-2.5mm}

\bb{slav3}D.A.~Slavnov.  Theor. Math. Phys., 136 (2003) 1273
\vspace {-2.5mm}

\bb{nev}   J.~Neveu.  Bases math\'ematiques du calcul des
probilit\'es,  Paris: Masson (1968) \vspace{-2.5mm}

\bb{epr}  A.~Einstein,  B.~Podolsky, and  N.~Rosen.  Phys. Rev.,
47 (1935) 777.  \vspace{-2.5mm}

\bb{dix}  J.~Dixmier. Les $C^*$ alg\`ebres et leurs
repr\'esentations,  Paris: Gauthier -- Villars \'Editeur 1969.
\vspace{-2.5mm}

\bb{slav4}  D.A.~Slavnov.  ArXiv:quant-ph/0101139 \vspace{-2.5mm}

\bb{bog} N.N.~Bogoliubov and D.V.~Shirkov. Introduction to Quantum
Fields, Wiley, New York 1980. \vspace{-2.5mm}

\bb{bell}J.S.~Bell.   Physics (Long Island City, N.Y.  1 (1965)
195 \vspace{-2.5mm}

\bb{chsh}J.F.~Clauser,  M.A.~Horn,  A.~Shimony,  R.A.~Holt. Phys.
Rev. Lett.,   23  (1969) 880 \vspace{-2.5mm}

\bb{slav5}D.A.~Slavnov.  Theor. Math. Phys., 106 (1996) 220.
\vspace {-2.5mm}

\bb{nam} M.~Namiki,  S.~Pascazio.  Phys. Rep.   232 (1993)  301.
\vspace{-2.5mm}

\bb{wheel}  J.A.~Wheeler.  Mathematical Founation of Quantum
Theory, New  York: Academic Press 1978 \vspace{-2.5mm}

 \bb{ton}  A.~Tonomura.  Phys. Today,   41  (1990) 22. \vspace{-2.5mm}

\bb{bohm}  D.~Bohm.  Phys. Rev.,   85  (1952) 166

 \end{thebibliography}

\end{document}